# Reliability Analysis to overcome Black Hole Attack in Wireless Sensor Network

Deepali Virmani [1], Ankita Soni [2], Nikhil Batra [3]

Bhagwan Parshuram Institute of Technology, India

deepalivirmani@gmail.com[1], ankita24soni@gmail.com [2], nikhilbatra789@gmail.com [3]

*Abstract*—Wireless sensor networks are vulnerable to several attacks, one of them being the black hole attack. A black hole is a malicious node that attracts all the traffic in the network by advertising that it has the shortest path in the network. Once it receives the packet from other nodes, it drops all the packets causing loss of critical information. In this paper we propose a reliability analysis mechanism. The proposed reliability analysis scheme overcomes the shortcomings of existing cooperative black hole attack using AODV routing protocol. As soon as there is a path available for routing, its reliability is checked using the proposed scheme. The proposed reliability analysis scheme helps in achieving maximum reliability by minimizing the complexity of the system. The final path available after the reliability analysis using the proposed scheme will make the path secure enough to minimize the packet loss, end-to-end delay and the energy utilization of the network as well as maximize the network lifetime in return.

Keywords: Blackhole attack, AODV Protocol, Malicious Node, Reliability, WSN

## I. INTRODUCTION

A wireless sensor network consists of a large number of wireless sensor nodes which are randomly deployed in the network [1] and have the ability to communicate with other sensor nodes. The sensor nodes are mobile in nature and have self-organizing capability that makes them flexible for communication in areas that have geographical and terrestrial constraints such as disaster areas and battle fields [2, 3]. Sensor nodes arrange themselves dynamically to create route among them to form a wireless network on the fly [3]. Recent research on wireless sensor network shows that they are more vulnerable to attacks than static networks. Therefore, any security solutions that are applicable for static routing network don't work well for wireless sensor network.

Wireless sensor network requires much stronger and effective security methods to confront the attacks caused by the malicious nodes in the network. Some of the attacks caused by the malicious nodes are Black hole attack, worm hole attack, hello flood attack, gray hole attack, denial of service[4,5] etc.

In this paper, we focus on the cooperative black hole attack. A black hole attack is an active attack in which a compromised node consumes all the data of the network. A black hole node falsely replies to all the route request packets during the route setup to the destination[6]. Once it receives the packets, it drops all of them leading to loss of information. A cooperative black hole attack consists of many such compromised nodes that work together and cause serious damage to the whole network.

In this paper, we try to overcome the limitations of an existing algorithm that works against the cooperative black hole attack [3]. We propose an algorithm that measures the reliability of every path that is established between the source and the destination. By analysing the reliability of every route or path, we make the network more secure and reliable for communication. We show, via simulations that our proposed algorithm works better in comparison to the existing algorithm. We also evaluate, via simulations, our proposed algorithm and compare it with the existing solution in terms of reliability, packet loss and end-to-end delay. The analysis shows that our proposed algorithm is much more effective than the existing solutions in terms of reliability and thus, it makes the network security stronger and immune to attacks.

The rest of the paper is organized as follows. Section 2 gives the related works that have been carried out in the area of black hole attack. Section 3 explains about the proposed algorithm in detail. Section 4 provides the simulation results. Section 5 concludes the paper with a brief overview of the future work.

## II. RELATED WORK

Many researchers have proposed different mechanisms to prevent black hole attack. Some of the works existing works are as follows.

In [2], Virmani et al proposed an algorithm to detect malicious nodes in the network using Selective Repeat ARQ in watchdog. In this mechanism, a node X monitors the transmission that occurs between the source and the destination. If any node misbehaves, it is detected by this monitoring node X and report about the maliciousness of a node is given to the source.

Seong et al [7] proposed two solutions to prevent black hole attack. In the first solution, the source node finds more than one path to the destination i.e. redundant routes and then identifies which is a safe route and which is unsafe and contains malicious nodes. In the second solution, each node maintains two extra tables:  one for last-packet-sequence-numbers for the last packet sent to every node and the other for last packets during setting up the network. During the RREP phase, the destination node must include the sequence number of the last packet received from the source. When the source receives the RREP, it compares the last sequence number with the value saved in its table. If the two values match, the transmission will take place. Otherwise the node that replies is a malicious node.

Sun et al [8] proposed a neighbourhood based methodology to detect the black hole attack. It determines the neighbour set for each node and compares them to determine whether a black hole attack occurs in the network. It establishes a path to the true destination in order to minimize the impact of the black hole attack.

Medadian et al [9] proposed an algorithm to combat black hole attack in AODV routing protocol. In this algorithm, a number of rules are established to check the honesty of a node. The activity of a node is judged by its neighbour. Every neighbour sends its opinion about a particular node to the source. Based on these opinions, the source decides if the replier is a malicious node.

In [3], Hesiri et al proposed an algorithm to detect cooperative black hole attack in AODV routing protocol (Figure 1). It consists of a DRI table with from and through entries. This table keeps track of whether or not the node did data transfers with its neighbours. To check the reliability of a node, the source asks the NHN (next hop neighbour) for the DRI entry of the node. Based on this information, source determines the reliability of a node. But this algorithm suffers from several limitations.

1) Large overhead: A lot of computation is required in this algorithm because for checking the reliability of the network 3 piece information is being brought from the NHN.
2) Collusion: Two nodes in the path (intermediate and next hop neighbour) can easily deceive the source node by sending the false information.
3) Endangered Information: When information is brought from the NHN then the information passes through the intermediate node only. If it is malicious it can easily tamper the information.
4) Go NO-GO Signal: Another problem is in the DRI Table. The flags stored are not enough to judge that the path is reliable or not.
5) Memory wastage: Each node stores the information about all the nodes which falls in its transmitting path.

## III. PROPOSED MODEL

In our proposed model, we conglomerate AODV protocol with reliability analysis to detect malicious nodes. The limitations mentioned above are overcome by reliability analysis scheme. We discuss various aspects of the model in the sub-sections below.

### A. Data Routing Information (DRI) Table

Similar to the [3] we use a DRI table in our proposed scheme. But we maintain the number of packets sent and received in our table instead of maintaining flags. A DRI table determines whether a node has forwarded data through its neighbours and whether it has received data from its neighbours. Each neighbour has a DRI entry which consists of the number of packets sent to a node and the number of packets received from that particular node. The DRI table is updated when a neighbour sends data packet to a node or receives data packets from the node.

Based on the entries (No. of Packets Sent, No. of Packets Received) in the table, we calculate the reliability ratio of the route consisting of the neighbours of the node.

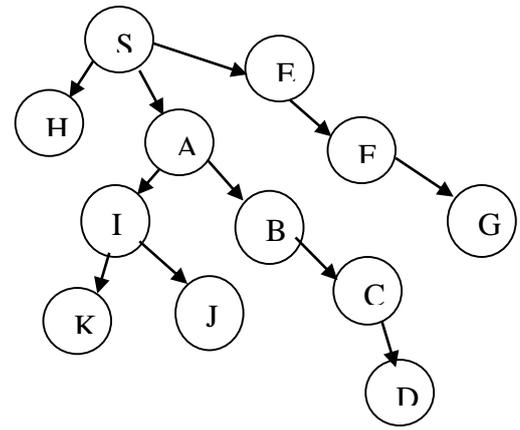

Figure 1: AODV Flooding

$$Reliability\ Ratio = \frac{No.of\ Packets\ Sent}{No.of\ Packets\ Received} \quad (1)$$

DRI Table is shown in Table 1. The table includes three columns; Packet ID, Number of Packets Sent, and Number of Packets Received.

**Table 1: DRI Table**

| Packet ID | No. of Packets Sent | No. of Packets Received |
|---|---|---|
| | | |
| | | |

### B. REL Packet

This is a special packet which is sent after the route has been discovered. REL packet keeps track of the reliability of each node.

$$REL = REL + Reliability\ Ratio \quad (2)$$

When the DRI entry for a node is requested from the NHN and a reply is received, it is checked if this reply packet contains the same information that the source node has for this NHN. If the two entries match, the REL (reliability) is updated by adding the reliability ratio of the NHN to the existing REL. This process continues for each NHN of the source. The REL for every NHN is returned to the source node on the basis of which, the source node selects the path with maximum reliability. Given below is the format of the REL Packet (Figure 2).

| Source | Destination | Next Hop neighbour | REL |
|---|---|---|---|

Figure 2: REL Packet

### C. Routing Table

In our proposed scheme, we introduce two new columns in the Routing table -Reliability Count and Hop Count. Reliability Count consists of submission of reliability of all the nodes. With the help of reliability column we are able to compare the reliability of different paths. Hop count determines the no. of hops between the source and the destination (See Table 2).

#### Table 2: Routing Table

| Source | Destination | Next Hop neighbour | Reliability Count | Hop Count |
|---|---|---|---|---|
|  |  |  |  |  |
|  |  |  |  |  |

*D. Scheme*

The first step consists of route discovery using AODV Protocol. When a node has to send packets to another node, it checks in its routing table whether a route to that destination exists. If the destination exists, then the source will ping the destination. After the source pings the destination, if a reply packet comes from the destination then the path to the destination is already established and so, a list of Intermediate Nodes (IN) is generated for the source. In case the reply doesn't come back from the destination within a specified time or the route to the destination doesn't exist, then RREQ (Route Request) packets are broadcast by the source to discover a route to the destination. Nodes that receive this RREQ packet either reply with a RREP packet or broadcast the RREQ packet further in the network. A list of IN is generated after receiving all the RREP packets.

Now let us consider a path from the source node S to the destination node D where A, B, C are the intermediate nodes (See Figure 3). As mentioned before, a black hole node is the one that consumes all the packets and transmits none. So, it's guaranteed that the Source node S that is transmitting packets to other node, cannot be a black hole node and that is why we consider S a reliable node.

S requests for its DRI entry from its NHN i.e. node A. At the same time, the feedback timer ($t_F$) is initialized to some real-time delay. If the NHN replies within the specified time, then the reply is accepted. If the feedback timer expires and no reply has been received, then another request will be sent to the NHN. This process repeats till the counter for reply ($C_R$) reaches its threshold value before the reply is received. Once the CR reaches its threshold value, source node is notified and the counter for malicious node ($C_M$) is incremented by one.

When S receives the DRI entry from node A, it cross checks the entry to the one maintained in its buffer for the NHN's. If the two entries do not match, then REL is set to ZERO and CM is incremented. If CM exceeds the threshold value, the path is declared ass UNTRUSTED.

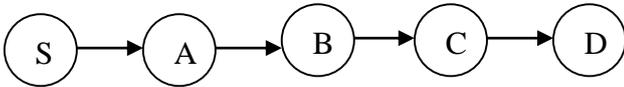

Figure 3: Path Established

But if the two entries match, then Reliability Ratio is calculated and added to the existing REL. Then S checks whether its NHN is the destination or not. If it is then no need to send the REL packet further but if it is not REL is forwarded to the NHN and the whole process is repeated. In this way if the entry keeps on matching REL will keep on summing up in the REL packet. After the packet reaches its destination then it is returned back to the source node where the entry for reliability is made. Finally the average is taken up for all paths by applying the mean route reliability (MRR) and the path will highest reliability is chosen.

1. Start
2. $C_m=0$, $C_R=0$ and $t_F=T1$
3. **IF** Destination in Routing table **THEN**
    3.1. Ping Destination
    3.2. **IF** Reply Received **THEN**
        3.2.1. Go to 6
            **ELSE**
        3.2.2. Go to 4
    **ELSE**
4. RREQ
5. RREP
6. List of IN is generated
7. **IF** NHN is Destination **THEN**
    7.1. Go to 8
        **ELSE**
    7.2. Request DRI Entry from NHN for current Node
    7.3. **IF** $t_F==0$ **THEN**
        7.3.1. $C_R = C_R + 1$
        7.3.2. **IF** $C_R > k$ **THEN**
            7.3.2.1. $C_R=0$
            7.3.2.2. Go to 7.2
                **ELSE**
            7.3.2.3. Go to 7.2
            **ELSE**
        7.3.3. **IF** Reply Received **THEN**
            7.3.3.1. **IF** the entry is Matched **THEN**
            7.3.3.2. REL = REL + Reliability Ratio
            7.3.3.3. REL packet is forwarded to NHN
            7.3.3.4. Go to 7
                **ELSE**
            7.3.3.5. $C_m=C_m+1$
            7.3.3.6. REL=0
            7.3.3.7. **IF** $C_m>k$ **THEN**
                7.3.3.7.1. Path is untrusted
                7.3.3.7.2. Go to 11
                    **ELSE**
                7.3.3.7.3. Go to 8
8. REL packet sent back to Source
9. MRR is calculated
10. Path with maximum reliability is selected
11. END

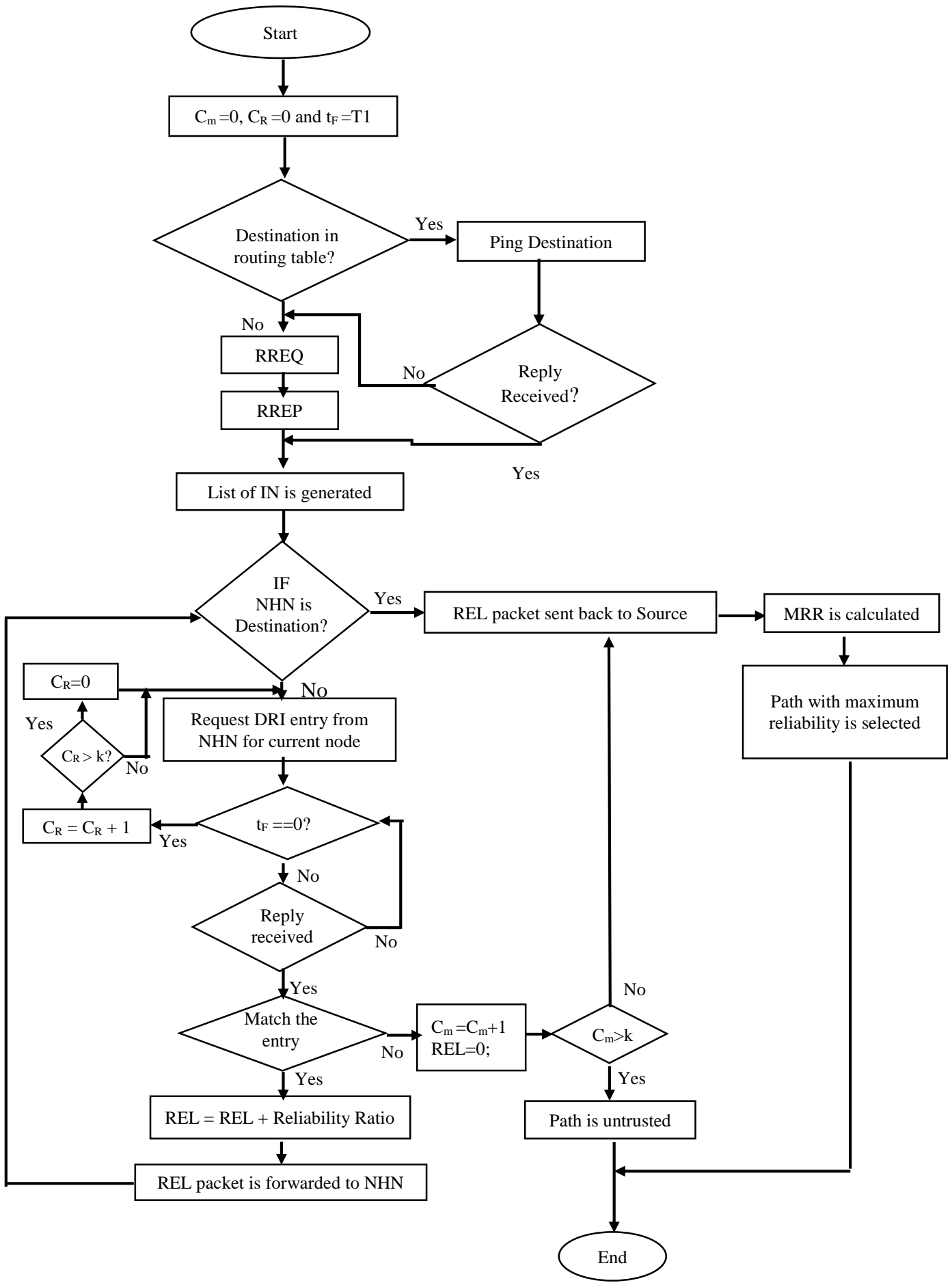

Figure 4: Proposed Flowchart

*E. Algorithm*

The flow chart for the proposed algorithm is shown in Figure 4. There are two counters maintained in the proposed model:

*1) $t_F$ :* It is a timer for feedback which ensures that the entries are received within given time frame.

*2) $C_m$:* It is counter for detecting malicious node. When it hits it threshold then we assume that the node is malicious.

*3) $C_R$:* This is a counter which is incremented every time when the $t_F$ is incremented. When this counter hits it threshold then $C_m$ is incremented by one.

## IV. RESULTS

In this paper, we consider four performance metrics to evaluate our proposed solution and to compare it against other existing solutions to prevent black hole attack. Since a black hole node drops the packets, there is a considerable loss of packets in the network. So packet loss is a performance metric in our analysis. Packet loss affects the throughput ratio of the network and so, we consider this metric as one of the performance metrics. We also take into account the end-to-end delay as another performance metric. Mean Route Reliability is one of the most important performance metric that we have considered in this paper and we show that the route selected by our proposed scheme is the most reliable of all.

The formulae for the four metrics is given as follows:
1) Throughput Ratio
   The throughput is the number of bytes transmitted or received per second. The throughput ratio, denoted by η, is calculated as follows:

$$\eta = \frac{\sum_{i=1}^{k} \eta_i^r}{\sum_{i=1}^{k} \eta_i^s} \times 100 \quad (3)$$

Where $\eta_i^r$ is the average receiving throughput for the ith application, $\eta_i^s$ is the average sending throughput for the ith application, and k is the number of applications[3].

2) Packet Loss Percentage
   Data packet loss rate, L is calculated as follows:

$$L = \frac{\sum_{i=1}^{k}(P_i^s - P_i^r)}{\sum_{i=1}^{k} P_i^s} \times 100 \quad (4)$$

Where $P_i^s$ and $P_i^r$ are the number of data packets sent by the sender and the number of data packets received by the receiver, respectively for the ith application, and k is the number of applications[3].

3) End-to-End Delay
   Average end-to-end delay of the data packets, denoted by E, is calculated as follows:

$$E = \frac{\sum_{i=1}^{k} e_i}{k} \quad (5)$$

where $e_i$ is the average end-to-end delay of data packets of ith application and k is the number of applications[3].

4) Mean Route Reliability

$$MRR = \frac{Reliability\ Ratio}{Hopcount} \quad (6)$$

Where Reliability Ratio and Hop count have already been discussed above.

For evaluation of our proposed scheme and comparison of our proposed scheme with the existing solution proposed by Hesiri in [3], we implement both the solutions and show simulation results using MATLAB. We carry out 100 simulation runs to get the appropriate results. We measure throughput, packet loss, end-to-end delay against the number of black hole nodes in the network and also measure reliability as a period of time.

Figure 5 shows throughput with respect to the number of black hole nodes. We consider 10 black hole nodes in our network. The simulation results show that the Hesiri [3] protocol has 65% throughput. But our proposed scheme provides a throughput of 76%. This depicts an increase of 16.92% throughput with our proposed scheme which shows that our proposed reliability scheme is better in terms of throughput than the existing solution proposed by Hesiri [3].

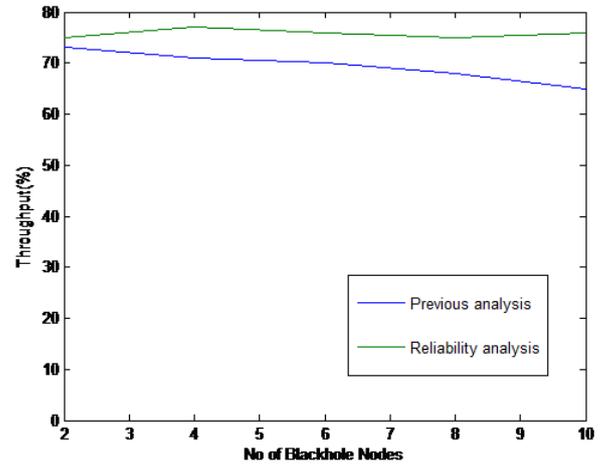

Figure 5: Throughput **vs. No of** Blackhole Nodes

Figure 6 shows the packet loss percentage with respect to the number of black hole nodes. The simulation results show that the existing protocol by Hesiri depicts a packet loss percentage of 50% whereas our proposed scheme shows a packet loss percentage of 20% only. Thus, the packet loss percentage shows a decrease of 60% in our proposed scheme.

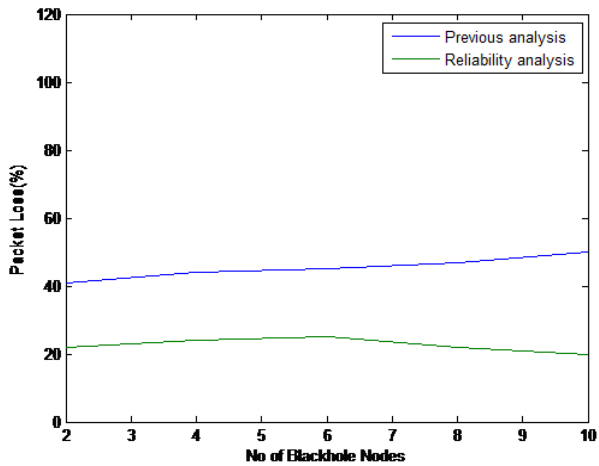

Figure 6: Packet Loss vs No of Blackhole Nodes

Figure 7 shows the end-to-end delay with respect to the number of black hole nodes. The simulation results show that our proposed scheme shows an end-to-end delay of 0.065 whereas the existing solution shows an end-to-end delay of 0.092. Thus, the decrease of end-to-end delay in our proposed scheme is 29.34%.

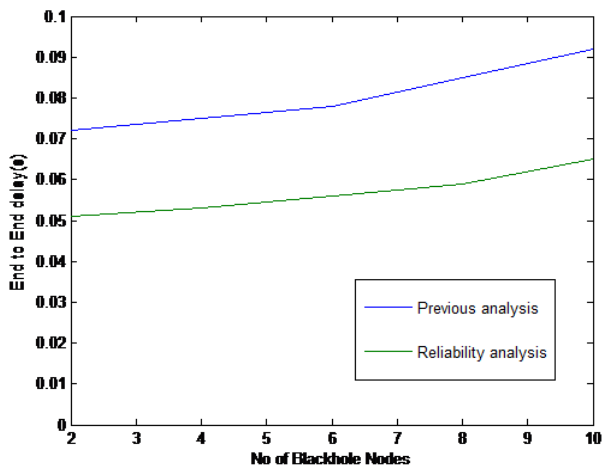

Figure 7: End to End Delay vs No. of Blackhole nodes

Figure 8 shows the reliability as a period of time. The simulation results show that our proposed algorithm has a reliability of 99% whereas the existing solution has a reliability of 60%. The increase in reliability is 66% for our proposed scheme.

All the above simulation results illustrates that our proposed algorithm works better than the existing solutions for the removal of black hole attack in the network. scheme.

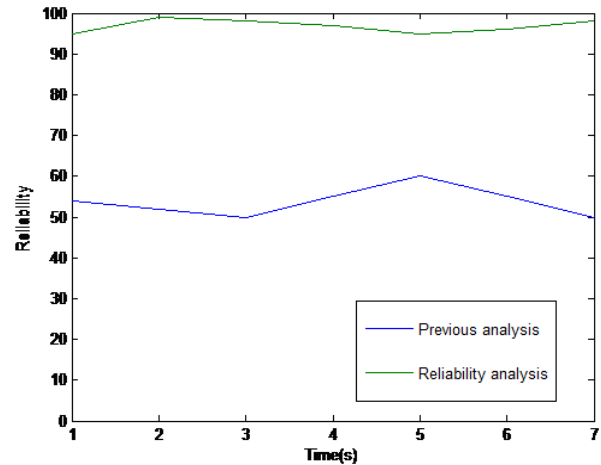

Figure 8: Reliability vs Time

## V. CONCLUSION

Wireless sensor network are susceptible to black hole attack that severely affects the whole network. To counter black hole attack, we proposed an algorithm that prevents this attack by detecting the black hole nodes. Our simulation results show that our proposed scheme improves the network security by establishing a secure route which has maximum reliability. We also show, via simulation results, that our proposed scheme is better than existing schemes for preventing black hole attack in terms of reliability, throughput, end-to-end delay and packet loss. Thus, our proposed scheme works successfully to counter black hole attack. In future, we aim to focus on other type of attacks that affect the network and try to develop methods to eliminate them from the network.